\documentclass[10pt,journal,compsoc]{IEEEtran}

\usepackage{lipsum}
\usepackage{fancyhdr}
\usepackage{xcolor}
\usepackage{graphicx}
\usepackage{subfigure}
\usepackage{caption}
\usepackage{float}
\usepackage{amsmath}
\usepackage{breqn}
\usepackage{empheq}
\usepackage{url}
\usepackage{epstopdf}
\usepackage{array}
\usepackage{booktabs}
\ifCLASSOPTIONcompsoc
  \usepackage[nocompress]{cite}
\else
  \usepackage{cite}
\fi
\ifCLASSINFOpdf
\else
\fi

\begin{document} 

\title{Localizable Button Click Rendering via Active Lateral Force Feedback}

\author{Heng~Xu$^{1}$,~\IEEEmembership{Member,~IEEE,}
        Roberta~L.~Klatzky$^{2}$, ~\IEEEmembership{Fellow,~IEEE,}
        Michael~A.~Peshkin$^{1}$,~\IEEEmembership{Senior~Member, IEEE}
        and~J.~Edward~Colgate$^{1}$,~\IEEEmembership{Fellow,~IEEE}
\IEEEcompsocitemizethanks{\IEEEcompsocthanksitem $^{1}$Department of Mechanical Engineering, Northwestern University, Evanston, IL, USA 60208. E-mail: hengxu@u.northwestern.edu, \{peshkin,colgate\}@northwestern.edu.
\IEEEcompsocthanksitem $^{2}$Department of Psychology, Carnegie Mellon University, Pittsburgh, PA, USA 15213. E-mail: klatzky@cmu.edu.}
\thanks{}}

\IEEEtitleabstractindextext{%
\begin{abstract}
We have developed a novel button click rendering mechanism based on active lateral force feedback. The effect can be localized because electroadhesion between a finger and a surface can be localized. Psychophysical experiments were conducted to evaluate the quality of a rendered button click, which subjects judged to be acceptable. Both the experiment results and the subjects' comments confirm that this button click rendering mechanism has the ability to generate a range of realistic button click sensations that could match subjects' different preferences. We can thus generate a button click on a flat surface without macroscopic motion of the surface in the lateral or normal direction, and we can localize this haptic effect to an individual finger. 

\end{abstract}

}

\maketitle
\thispagestyle{fancy}
\IEEEdisplaynontitleabstractindextext

\IEEEpeerreviewmaketitle

\IEEEraisesectionheading{\section{Introduction}\label{sec:introduction}}

\IEEEPARstart{P}{rofessional} tablets and two-in-ones (such as the Microsoft Surface) are growing in popularity at the expense of traditional laptop computers. Laptops, however, offer a key advantage: keyboards that enable touch-typing in which at least some of the fingertips rest on the keys. Keys are activated by force rather than by contact. At present, touch-typing remains one of the highest-bandwidth means of communicating information from a human to a computer. Keyboards, however, take up space that often goes unused; as such, an exciting development would be a touchscreen keyboard that supported touch-typing. Requirements for such a device would include localized pressure sensing, tactile feedback, and mechanical simplicity (e.g., few moving parts).

Many researchers have studied how to render a button click sensation, some employing vibration stimuli in the normal direction and some in the lateral direction. For example, Fukumoto et al. and Chen et al. used vibrations having a sinusoidal waveform in the normal direction to simulate the click sensation \cite{fukumoto2001active,chen2011design}. A difficulty with this method, however, is that, unlike physical buttons which typically exhibit a single sharp transient \cite{weir2004haptic}, repeated oscillations (more than three) result ``in an eerie sensation of something alive'' \cite{chen2011design}. Zoller et al. overcame this problem by using a thin electromagnetic actuator module on a capacitive touchscreen to provide a single sharp transient  \cite{zoller2012novel}. A remaining limitation of normal direction methods, however, is that it is difficult to localize the click sensation to an individual finger.

It is also possible to render clicks with vibrations in the lateral direction. For instance, a commercial force touch trackpad (MacBook Pro Retina 2015, Apple) employed an electromagnetic linear actuator to provide click feedback in the lateral direction \cite{kessler2015haptic}.  Gueorguiev et al. also used active lateral force feedback to simulate a click sensation \cite{gueorguiev2018travelling}. This method, however, was based on a traveling wave having a lower Q factor than actuation based on a standing wave, such that a bulky actuator was required to generate lateral forces for the click sensation. Also, Gueorguiev et al. reported that this method did not show an advantage over the ultrasonic friction modulation method on the subjective quality of the click sensation \cite{gueorguiev2018travelling}.

Monnoyer et al. and Tashiro et al. \cite{tashiro2009realization,monnoyrer2018perception} used ultrasonic vibrations to modulate the friction between the fingertip and surface. They showed that some people could feel a click sensation if a transition from high friction to low friction occurred as the finger pushed on the surface; however, the sensation depended on the impedance of each individual's fingertip \cite{monnoyrer2018perception}. More generally, it is difficult to generate strong haptic effects via friction modulation unless the finger and surface are sliding relative to one another in the lateral direction.

These lateral excitation methods described so far, however, make no attempt to localize the click sensation: all fingers touching the surface feel the same click. To address this limitation, Hudin and colleagues proposed a time-reversal wave focusing method that could be used to create high amplitude ultrasonic vibration at localized points on the surface \cite{hudin2015localized}. A finger placed over one of these points would be ``ejected'' (thrown off the surface), which was easily perceived. This method, although elegant, provided very little control over the waveform applied to the finger, and also produced an undesirable audible artifact. In subsequent work, Hudin proposed another method called non-radiating ultrasonic vibrations \cite{hudin2017local} and demonstrated independent control of the ultrasonic vibration at different positions on a surface by using two piezoelectric actuators. Even though the non-radiating ultrasonic vibration method is able to localize friction modulation, the vibration fields are wholly dependent on the position of the actuators. Thus, it is difficult for this method to localize friction modulation at more than a few points on the surface.

Extending our previous work (the UltraShiver \cite{xu2019ultrashiver,xu2019localized}), this paper proposes a new method for rendering a button click and localizing this effect by localizing electroadhesion between a finger and a surface, thus providing a promising method for simulating a touch-typing keyboard. In addition to localized pressure sensing, requirements for touch-typing include click rendering as well as localized control of this rendering. We report on three experiments that were conducted to investigate the ability of the UltraShiver to create and localize a button click sensation. Section \ref{localizedControl} demonstrates the ability of the UltraShiver to localize control of the active lateral force on the fingertip. Section \ref{buttonClick}, uses the force profile of pressing on the virtual button to show the robust control of the UltraShiver for the button click rendering. Finally, psychophysical experiments are used to evaluate the quality of button click rendering and its relationship to stimulus parameters. Overall, the UltraShiver not only simulates the button click sensation but also localizes the effect, presenting a promising method for touch-typing keyboard rendering.

\section{Background of UltraShiver}
The new method presented here is based on a lateral force feedback device, the UltraShiver, which we presented in a previous study \cite{xu2019ultrashiver}. The UltraShiver consists of two piezoelectric actuators and a sheet of anodized aluminum (shown in Fig. \ref{fig_lateralDisplacement_setup}). The dimensions of the anodized aluminum are 84 mm x 60 mm x 1 mm. Two pieces of hard piezoceramic (SMPL60W5T03R112, Steminc and Martins Inc, Miami, FL, USA) are placed in the middle of the anodized aluminum. The lateral force generation of the UltraShiver depends on synchronization of in-plane ultrasonic oscillation and out-of-plane electroadhesive forces. The in-plane ultrasonic oscillation is due to the longitudinal resonance of the UltraShiver and is excited by two piezos. The out-of-plane electroadhesive force is controlled by an electric current applied between the fingertip and the surface of the anodized aluminum. Significantly, both the in-plane oscillation and the electroadhesive force are modulated at an ultrasonic frequency well beyond the tactile and audio bandwidths: neither one can, in isolation, be perceived.  When the two are present together, however, a non-zero average lateral force is produced.  This happens because the in-plane vibrations cause the direction of the friction force to oscillate, while the electroadhesion causes the magnitude of the friction force to oscillate at the same frequency.  Thus, it is possible for the average friction force to be higher in one direction than the other. By adjusting the phase between the ultrasonic oscillation and the electroadhesive force, the direction and magnitude of the lateral force can be controlled. According to \cite{xu2019ultrashiver}, the relation between active lateral force and the phase difference can be approximated by the following equation:
\begin{dmath}
  {F_{lf}} = {-\mu _k}\frac{{2A{\varepsilon _0}{\varepsilon _{gap}}}}{{d_{gap}^2}}\frac{{V_{ACgap}V_{DCgap}}}{\pi }\cos(\varphi)
\label{eq:1}
\end{dmath}
where $F_{lf}$, $\varphi$, $V_{ACgap}$, and $V_{DCgap}$ are the active lateral force, the phase between the ultrasonic oscillation and the electroadhesive voltage, the AC voltage and DC voltage across the gap between the fingertip and the surface, respectively. The parameters $\mu _k$, $A$, $\varepsilon _0$, $\varepsilon _{gap}$, and $d_{gap}$ are the kinetic friction coefficient, gross contact area,  permittivity of free space, and dielectric constant and thickness of the air gap between the fingertip and the surface, respectively.

\section{Localized Control of Lateral Force} \label{localizedControl}
As one requirement of touch-typing keyboard rendering, the ability of the UltraShiver to localize control of the lateral force was investigated in terms of both lateral force generation on the fingertip and vibration propagation between the fingers. Experiment 1 was designed to investigate the latter topic by attempting to avoid lateral force on one of two fingers, both of which were touching the surface. This raised the question of how to localize lateral force. In principle, the lateral force generation of the UltraShiver depends on the synchronization of the ultrasonic oscillation and the electroadhesion. Thus, it should not produce lateral force if either the ultrasonic oscillation or the electroadhesion is absent. Compared with localizing the ultrasonic oscillation, localizing electroadhesion is easier. For instance, the conductive layer of the surface could be etched into a grid with each section of the grid selectively connected to the electroadhesion high voltage source based on the finger action and the algorithm of haptic effects \cite{peshkin2014haptic,colgate2017touch}. In addition, the distance between each section of the grid can be designed for a keyboard scenario. Since a single sheet of anodized aluminum that had isotropic electrical conductivity was employed in the current prototype of the UltraShiver, an insulating anodized aluminum foil (More details are found in Fig. \ref{fig_lateralDisplacement_setup} and Section \ref{experiment1Setup}) was used to create electrical isolation between one finger and the surface and thereby localize the electroadhesive force to the other finger.

\subsection{Experiment 1}
\subsubsection{Experiment Setup} \label{experiment1Setup}
\begin{figure}[!htb]
  \centering
  \subfigure[Side view of the apparatus.]{ 
    \label{fig_lateralDisplacement_setup:a} 
    \includegraphics[width=0.45\textwidth]{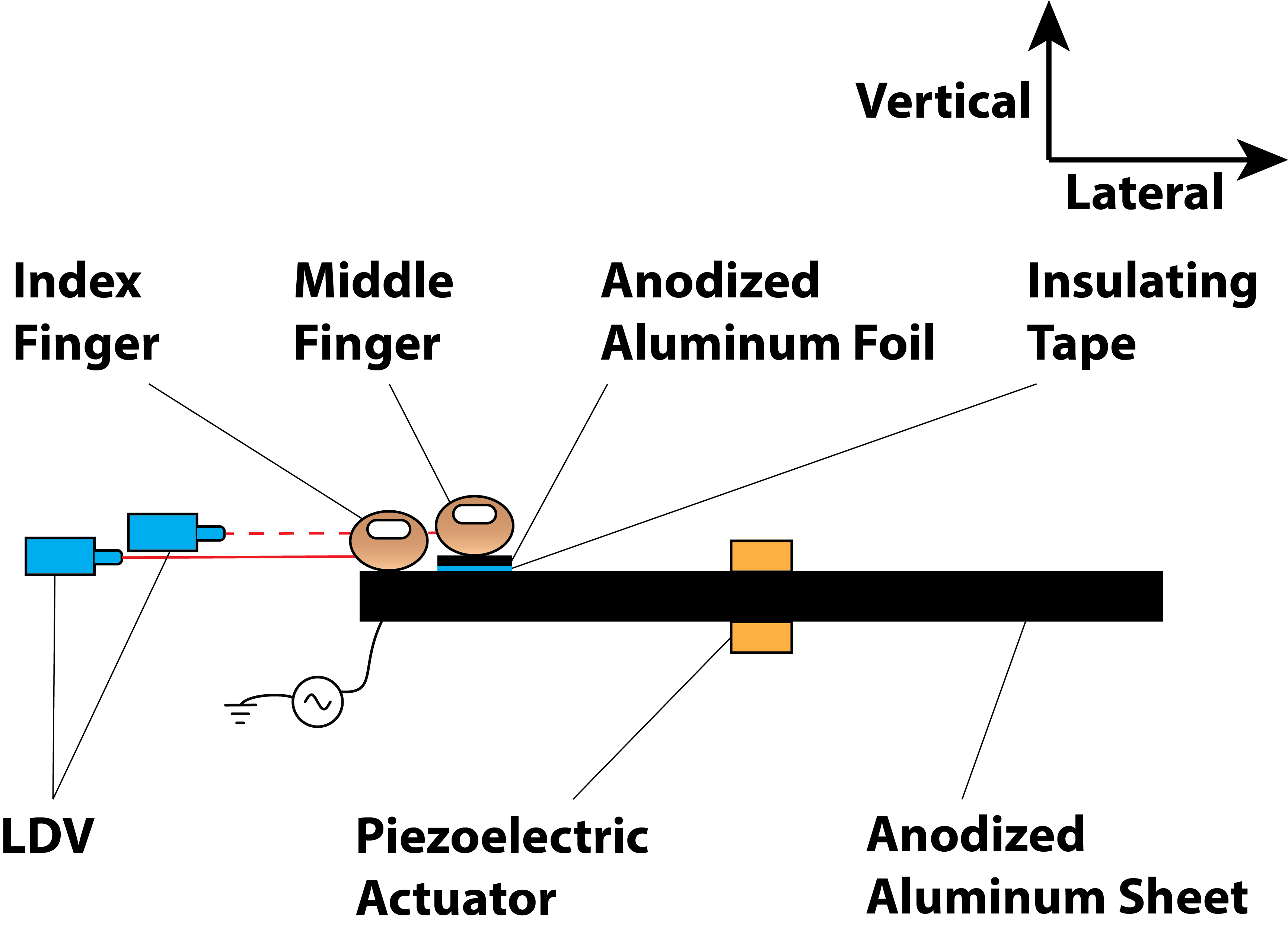}} 
  \hspace{0in} 
  \subfigure[Top view of the surface]{ 
    \label{fig_lateralDisplacement_setup:b} 
    \includegraphics[width=0.45\textwidth]{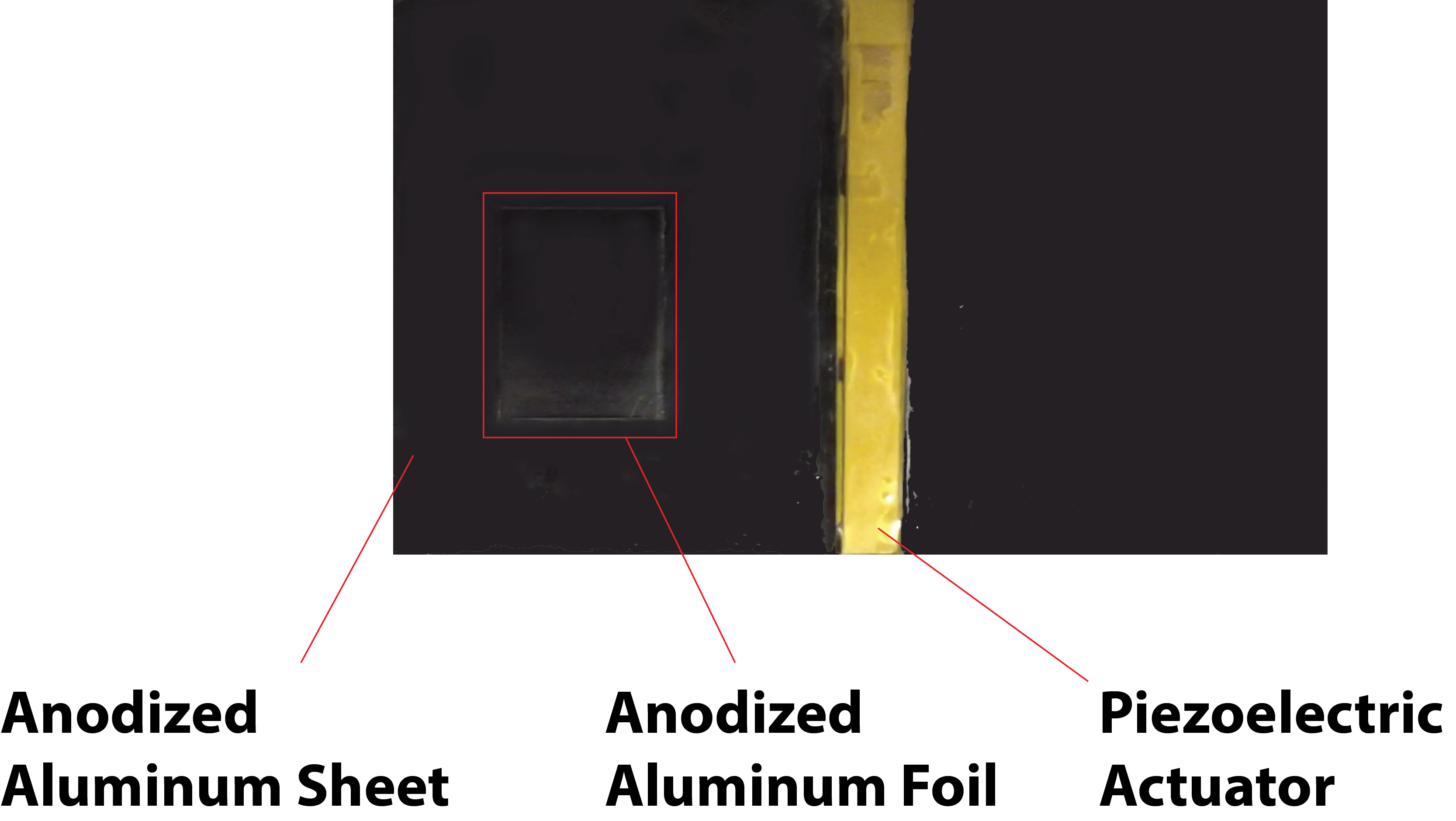}} 
  \caption{Experiment 1 setup.}
  \label{fig_lateralDisplacement_setup}
\end{figure}

Fig. \ref{fig_lateralDisplacement_setup} shows the setup in Experiment 1 (the mounting of the UltraShiver is not shown, but is the same as that in Fig. \ref{fig_forceMeasurement_setup}). The lateral displacements of the fingers were measured with a laser doppler vibrometer (LDV, IVS-500, Polytec, Inc). Measurements were made at points 1 mm above the surface. In Experiment 1, the lead author used the index finger and the middle finger of his dominant hand, which were electrically grounded, to press on the surface. The index finger contacted the anodized aluminum sheet, while the middle finger contacted the insulating foil. The foil was made from a 1 mm anodized aluminum sheet (18 mm x 15 mm) milled down to 60 $\mu m$ and backed with insulating tape. An epoxy adhesive (Acrylic Adhesive 3526 Light Cure, Loctite, Westlake, OH, USA) was used to connect the anodized aluminum foil, the insulating tape, and the anodized aluminum sheet. This provided an insulated patch that otherwise had the same contact characteristics as the main (electroadhesive) surface. 

The voltage of piezoelectric actuators and the electroadhesive current were controlled with a custom voltage amplifier and a custom high voltage source, respectively (more details were reported in \cite{shultz2018application,xu2019ultrashiver}). Since only one LDV was used in the experiment, it was repositioned in separate trials to measure the lateral displacement of each finger. All signals were recorded using a NI USB-6361 Multifunctional I/O Device with a 1 MHz sampling frequency.

\subsubsection{Experiment Protocol}
In this experiment, the frequencies of the electroadhesion voltage and the piezoelectric voltage were set to 29.99 kHz and 30 kHz, respectively, so that the lateral force on the fingertip varied at 10 Hz beat frequency. Lateral displacements of the fingers were measured with an LDV. The measurement points of the LDV were on the left sides of the fingers and close to the finger-surface contact (in Fig. \ref{fig_lateralDisplacement_setup}). Each measurement lasted 2 seconds and was repeated ten times. 

\subsection{Results}
\begin{figure}[!htb]
  \centering
  \subfigure[Lateral displacement at 10 Hz]{ 
    \label{fig:displacement_profile:a} 
    \includegraphics[width=0.45\textwidth]{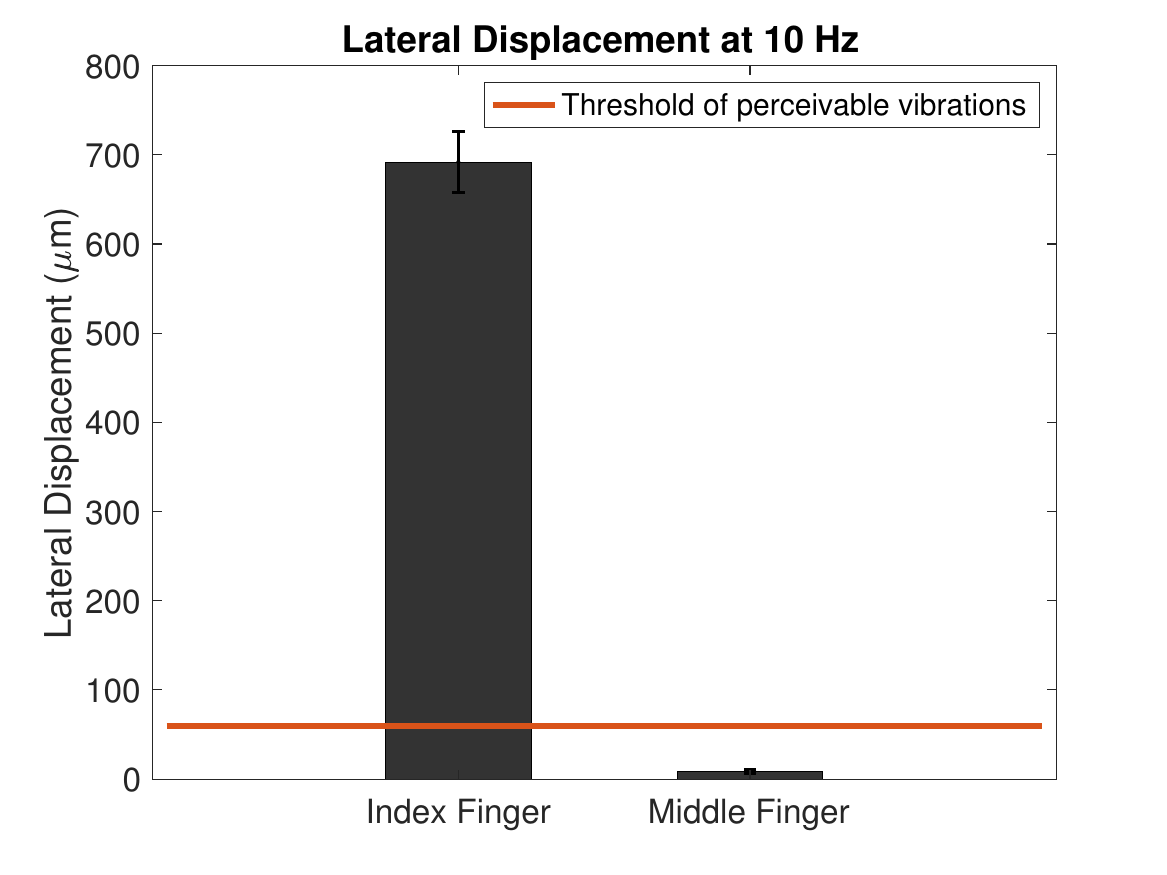}} 
  \hspace{0in} 
  \subfigure[Lateral displacement at 30 kHz]{ 
    \label{fig:displacement_profile:b} 
    \includegraphics[width=0.45\textwidth]{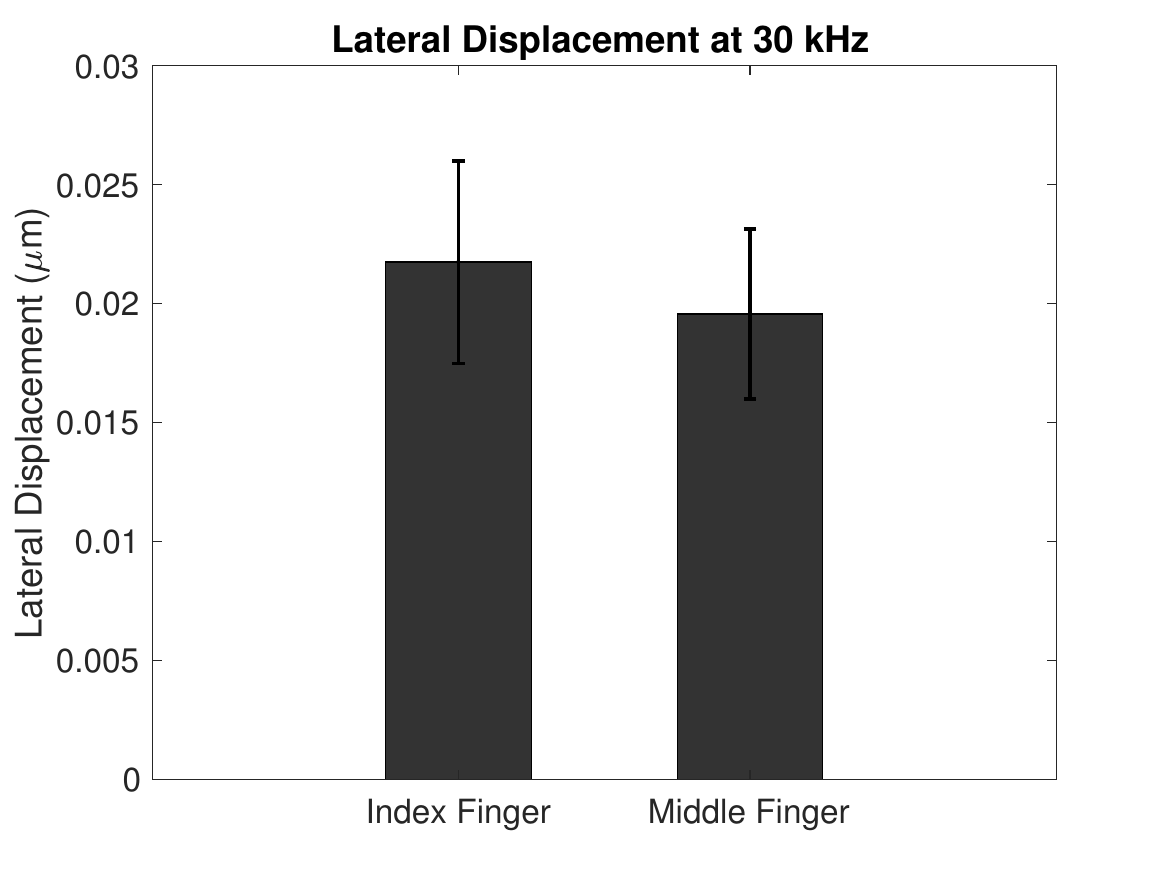}} 
  \caption{Lateral displacement envelope of the index finger and the middle finger when they touch the anodized aluminum sheet and the insulating anodized aluminum foil, respectively (in Fig. \ref{fig_lateralDisplacement_setup}). The frequencies of the electroadhesion voltage and the piezoelectric voltage were set to 29.99 kHz and 30 kHz, respectively, so that the lateral force on the fingertip varied at a 10 Hz beat frequency. Black bars represent the average displacement envelope; error bars represent the standard deviations over five trials. The red line is the threshold of vibration perception at 10 Hz.}
  \label{fig:displacement_profile}
\end{figure}

Figs. \ref{fig:displacement_profile:a} and \ref{fig:displacement_profile:b} show the lateral displacement envelopes of the index finger and the middle finger at 10 Hz and 30 kHz, respectively. In Fig. \ref{fig:displacement_profile:a}, the lateral displacement envelope of the index finger at 10 Hz is 691.9 $\pm$ 34.2 $\mu m$, which is significantly higher than that of the middle finger (8.6 $\pm$ 1.7 $\mu m$). 

While the human detection threshold of vibration in the lateral direction is not well documented in the literature, the threshold in the normal direction has been reported to have an amplitude of around 100 $\mu m$ at 10 Hz \cite{mountcastle1967neural,talbot1968sense,kandel2000principles}. It has been reported that mechanoreceptors are more sensitive to the vibration in the lateral direction with a detection threshold approximately 0.6 of that in the normal direction \cite{biggs2002tangential}. By this estimate, the amplitude threshold of perceivable vibration in the lateral direction at 10 Hz is around 60 $\mu m$ peak (as shown by the red line in Fig. \ref{fig:displacement_profile:a}), which is lower than the measured amplitude of the index finger and higher than that of the middle finger (see Fig. \ref{fig:displacement_profile:a}).

Fig. \ref{fig:displacement_profile:b} shows that the lateral displacement envelope of the index finger at 30 kHz is 0.022 $\pm$ 0.004 $\mu m$, which is close to that of the middle finger (0.020 $\pm$ 0.004 $\mu m$). Since these ultrasonic vibrations with a low amplitude are beyond the perceivable range of human cutaneous mechanoreceptors, subjects cannot detect them. 

These isolation results in Fig. \ref{fig:displacement_profile} show that the UltraShiver provides excellent localization of the active lateral force (More details are discussed in Section \ref{discussLocalized}).

\section{Button Click Rendering} \label{buttonClick}
In this section, two experiments are presented. Experiment 2 investigated the force generating capacities of the UltraShiver for button click rendering, while Experiment 3 studied the subjects' perceptual evaluation of the rendered clicks.

\subsection{Button Click Rendering Algorithm}
The button click rendering algorithm was based on modulation of the active lateral force on the fingertip which was achieved by adjusting the phase between the ultrasonic oscillation and the electroadhesive voltage (${0^{^\circ}}$: move finger to the left; ${180^{^\circ}}$: move finger to the right). Note that ultrasonic oscillations were operating at all times so as to avoid perceptual artifacts. When the pressing (normal) force crossed over a set threshold, a square-waveform lateral force was constructed and applied to the fingertip (see the command signal in Fig \ref{fig:force_profile:a}). The normal force threshold was 600 mN, a typical value taken from the measurement of a physical button (Logitech Keyboard K120). By varying the duration and duty cycle of the square waveform, the tactile characteristics of the rendered button click could be changed over a fairly broad range (more details in section \ref{Experiment 3}).

\subsection{Experiment 2: Force Profile Measurement of Button Click Rendering}
\subsubsection{Experiment Setup}
\begin{figure}[!htb]
\centering
\includegraphics[width = 0.45\textwidth]{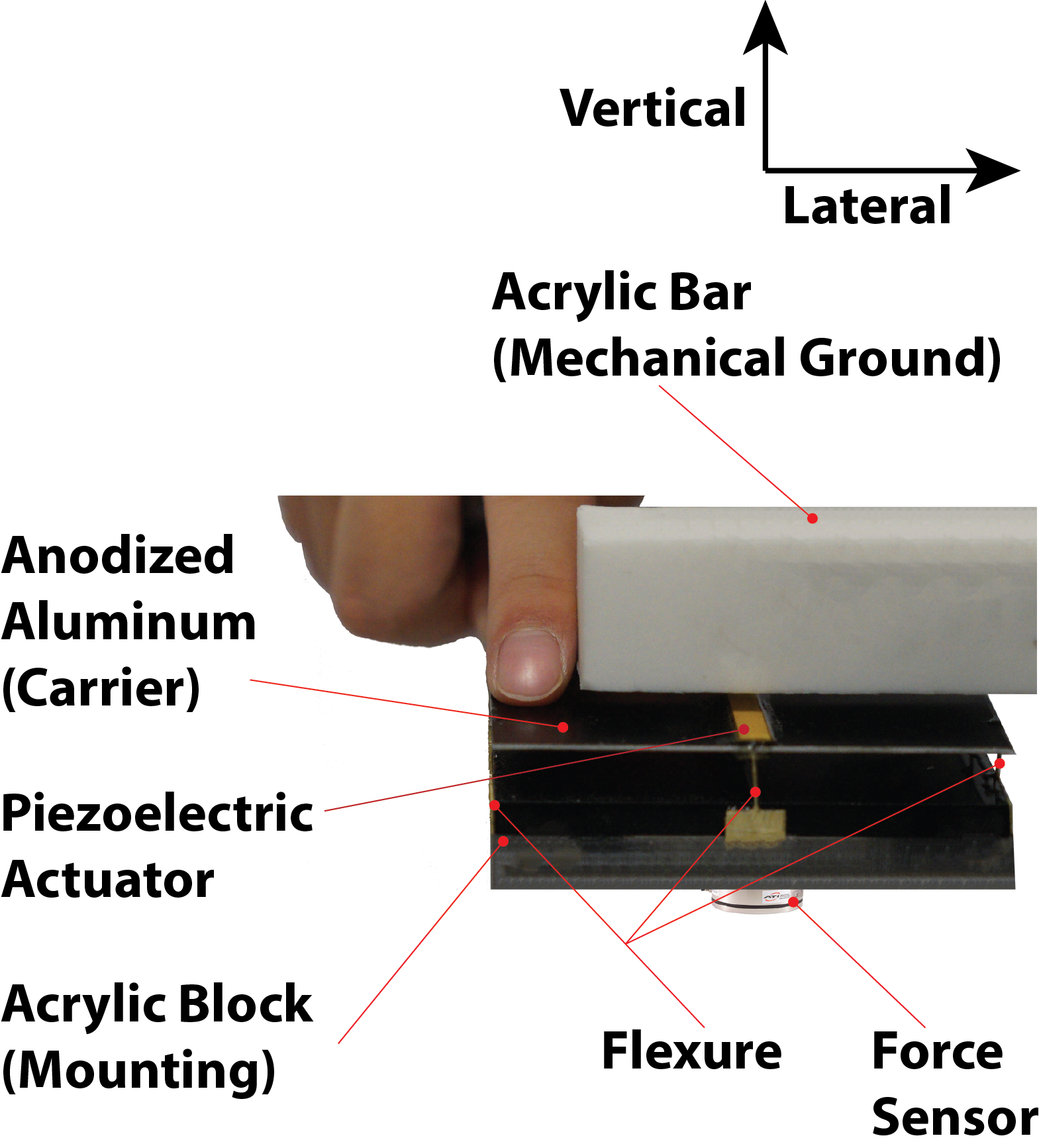}
\caption{Side view of the setup in Experiment 2.}
\label{fig_forceMeasurement_setup}
\end{figure}

Experiment 1 and Experiment 2 shared most parts of the setup (in Fig. \ref{fig_lateralDisplacement_setup} and Fig. \ref{fig_forceMeasurement_setup}), except for the number of involved fingers and finger movement constraints. The electrically grounded index finger of the dominant hand was used in Experiment 2. It was constrained to move only up and down while pressing against a white acrylic bar to mechanically ground the rest of the finger. As shown in Fig. \ref{fig_forceMeasurement_setup}, the UltraShiver was mounted to a black acrylic block with brass flexures (6 flexures, 10 mm tall x 1 mm x 0.3 mm). The acrylic block was fixed on a six-axis force sensor (ATI17 Nano load cell), which was used to measure the normal and lateral force. 

All signals were recorded using a NI USB-6361 Multifunctional I/O Device with a 300 kHz sampling frequency.

\subsubsection{Experiment Protocol}
During this experiment, the lead author pressed on the surface and then lifted up, as if pressing on a physical button. When the pressing force reached the set threshold (600 mN), the stimulus of button click rendering was applied to the finger. The lateral and normal forces were measured by the six-axis force sensor. There were fifteen trials in the experiment. During each trial, the subject pressed with the same amount of force, to the best of his ability. 

\subsection{Experiment 3: Perceptual Evaluation of Rendered Button Clicks} \label{Experiment 3}
In Experiment 3, perceptual experiments were designed to evaluate the quality and variety of the rendered button clicks that resulted in user acceptance. In our previous work \cite{xu2019localized}, we found that the shortest pulses of acceptable button click rendering were quite similar (below 26.4 milliseconds), which might be consistent with subjects' detecting only one event across the entire cycle (see the command signal in Fig \ref{fig:force_profile:a}). Thus, we proposed a hypothesis in \cite{xu2019localized}: the quality of button click rendering is related to the number of events perceived in the stimulus, and the detection of only one event is judged to be an acceptable button click. The verification of this hypothesis was also investigated in Experiment 3.

\subsubsection{Participants}
Ten subjects (20 to 30 years of age, one left-handed, four female) participated in this experiment. Seven of the subjects were naive to the purpose of the experiment and had no experience with surface haptics, while the other three subjects (subject 4, 5, and 9) were graduate students in the haptics group. The authors did not serve as subjects in this experiment. Subject participation was approved by the Northwestern Institutional Review Board, subjects gave informed consent, and subjects were paid for their time.

\subsubsection{Experiment Protocol}
\begin{figure}[!htb]
\centering
\includegraphics[width = 0.45\textwidth]{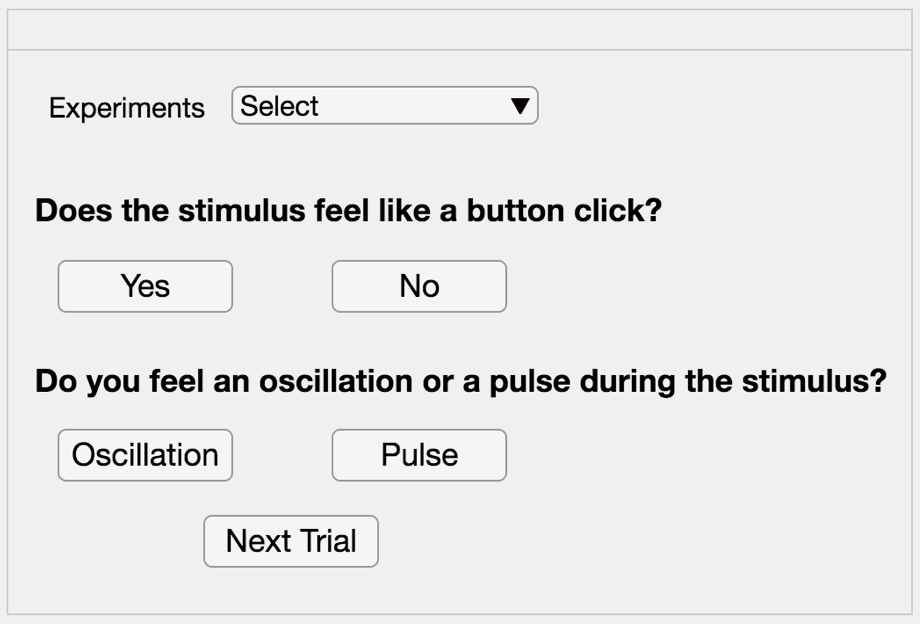}
\caption{GUI interface in Experiment 3: the first section.}
\label{fig_subjectJudge_interface}
\end{figure}

\begin{figure}[!htb]
\centering
\includegraphics[width = 0.45\textwidth]{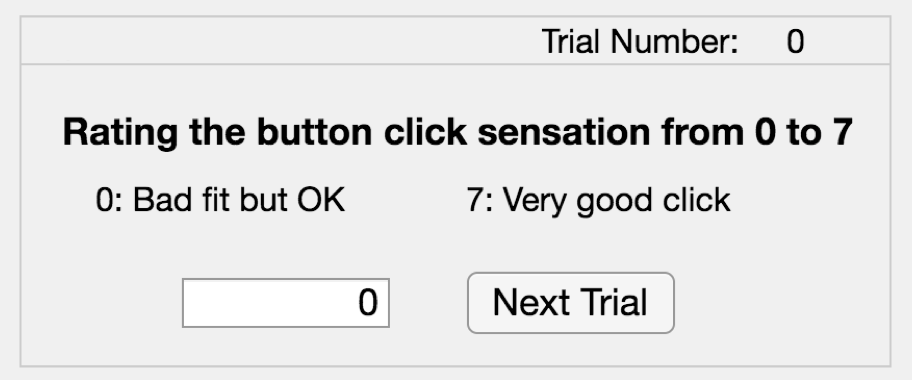}
\caption{GUI interface in Experiment 3: the second section.}
\label{fig_subjectRate_interface}
\end{figure}

Experiment 3 was conducted in two sections. The first section (shown in Fig. \ref{fig_subjectJudge_interface}) was designed to investigate the range of acceptable button click rendering and the relation between the acceptable button click rendering and the number of events perceived in the stimulus. The second section (shown in Fig. \ref{fig_subjectRate_interface}) was used to study the subjects' rating of their own acceptable rendering based on the results of the first section. There was a five-minute break between the two sections.

In the first section (shown in Fig. \ref{fig_subjectJudge_interface}), each stimulus consisted of one cycle of a rectangular waveform. The duty cycle and duration of the stimulus were adjusted to generate different button clicks (see the command signal in Fig. \ref{fig:force_profile:a}). The duty cycle was defined as a ratio of the duration with the positive lateral force to the total duration of the stimulus. The duty cycle was one of three levels: 5\%, 25\%, or 50\%. The duration was one of 26 levels, ranging from 1 millisecond to 251 milliseconds with equal intervals between levels.

There were six blocks in the first section. Each block employed a duty cycle from one of the three levels (5\%, 25\%, or 50\%), and swept through the durations along either an increasing or decreasing trajectory. The increasing trajectory meant that the duration started with the minimum value (1 millisecond) and increased to the maximum value (251 milliseconds) across 26 successive stimuli. The decreasing trajectory was the reverse. Thus, each stimulus with the same duration and duty cycle was presented twice, once in each sweep direction. Each block took around 5 minutes, and the total section lasted 30 - 40 minutes, including breaks.

During each block, subjects were asked to press on the surface with the index finger of their dominant hand, as if pressing on a physical button. They were further instructed to consistently press on the same contact patch area of the surface with a constant contact angle between the finger and the surface. Headphones playing pink noise were worn to cancel any sounds produced by the experimental platform. A yellow LED indicated whether the subject reached the normal force threshold of the button click. 

After each trial, subjects were asked two questions. The first question was whether the stimulus felt like an acceptable button click. The second question was whether the stimulus felt like an oscillation or a pulse. Subjects gave YES or NO verbal answers to the first question and OSCILLATION or PULSE verbal answers to the second question. These were recorded by the experimenter. Subjects made their judgment based on their own prior experience with buttons. For each subject, the first YES answer and the last YES answer were used to define the boundaries of the good-button range for each duty cycle. These boundaries were averaged over the increasing trajectory and the decreasing trajectory resulting in a minimum and a maximum duration for each duty cycle. 

In the second section (shown in Fig. \ref{fig_subjectRate_interface}), subjects were asked to rate the button click rendering within the acceptable ranges found in the first section. The rating was performed in three successive rounds. In the first round, five equally-spaced durations spanning from minimum acceptable to maximum acceptable were selected for each duty cycle. Each of these durations was randomly presented three times. Subjects were asked to press on the surface as they did in the first section and to rate the rendered button click sensation from 0 to 7. Responses were recorded by the experimenter. The second round proceeded similarly; however, the tested durations were separated by only 10 milliseconds and centered on the one receiving the highest rating in round one. In the third round, the tested durations were separated by only 5 milliseconds and centered on the one receiving the highest rating in round two. By way of example, suppose that a subject had, at the 25\% duty cycle, rated stimuli from 10 to 170 milliseconds as acceptable. Then the subject would be asked to rate durations of 10, 50, 90, 130, and 170 milliseconds in round one. If the subject awarded the 90 milliseconds duration the highest score, then she would be tested at 60, 70, 80, 90, 100, 110 and 120 milliseconds in round two (note that the set was truncated to lie strictly between the already-tested 50 and 130 millisecond durations). If the subject then awarded 70 milliseconds the highest score, she would be tested at 65, 70, and 75 milliseconds in round three. Although the total number of trials varied between subjects, this was an efficient procedure, typically lasting 10 - 15 minutes.

Before starting the experiment, subjects were asked to wash and dry their hands. They were exposed to samples of rendered button clicks and familiarized with the experimental platform.

\subsection{Results}
\subsubsection{Experiment 2: Force Profile Measurement of Button Click Rendering}
\begin{figure}[!htb]
  \centering
  \subfigure[Lateral force]{ 
    \label{fig:force_profile:a} 
    \includegraphics[width=0.45\textwidth]{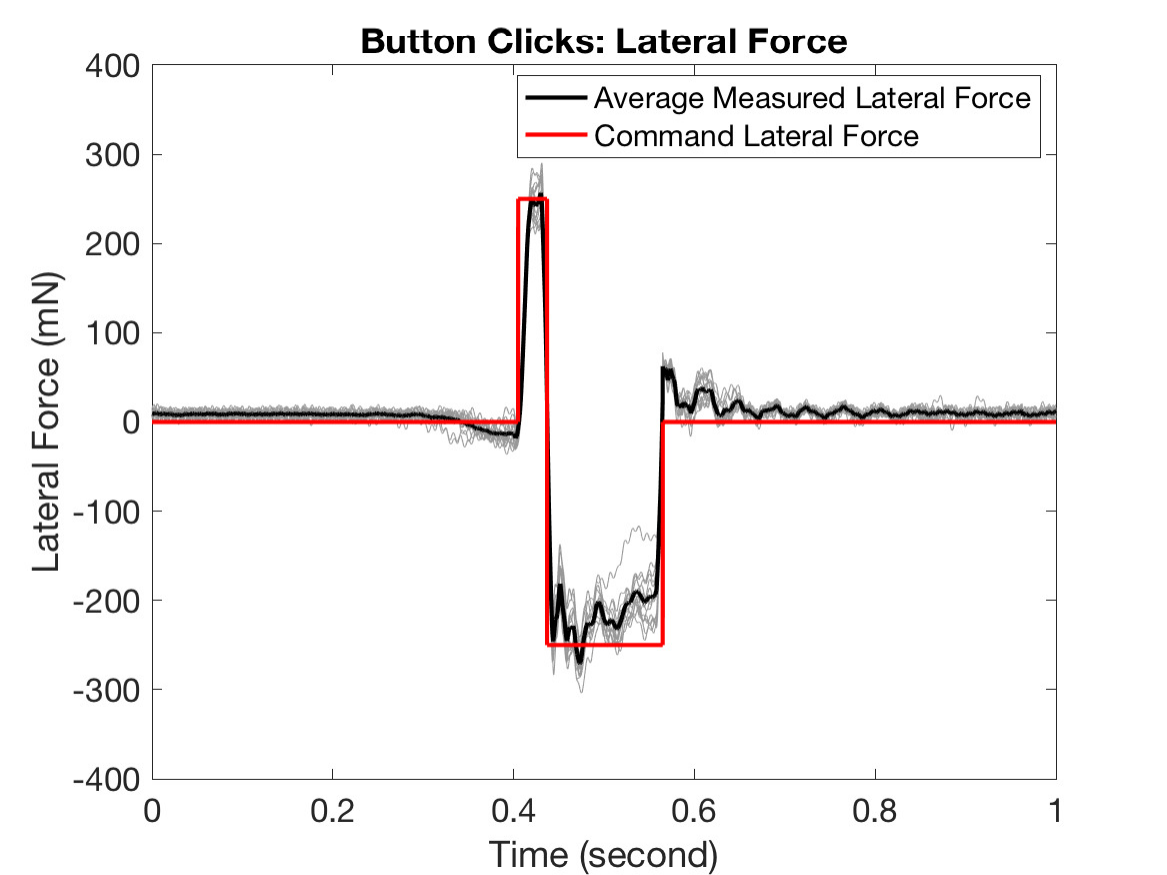}} 
  \hspace{0in} 
  \subfigure[Normal force]{ 
    \label{fig:force_profile:b} 
    \includegraphics[width=0.45\textwidth]{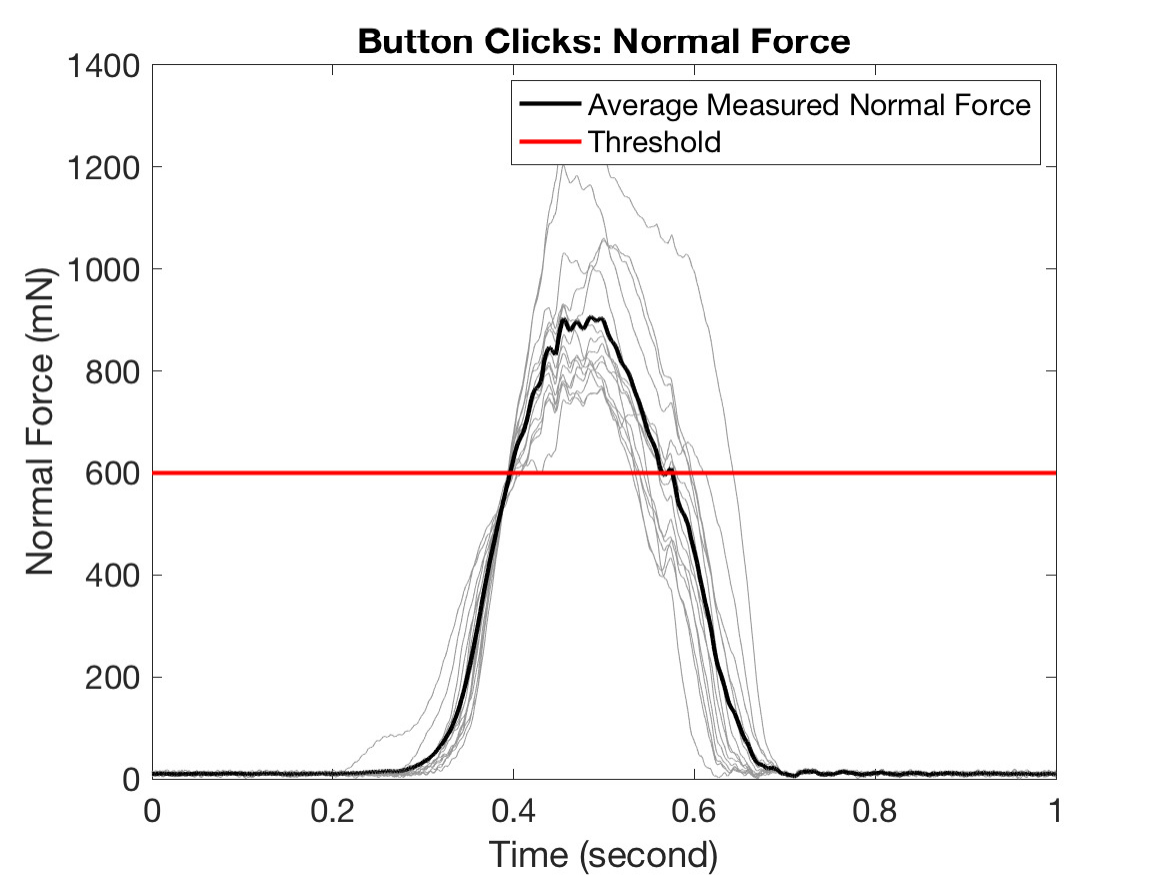}} 
  \caption{Lateral and normal force measurement during button click rendering. The gray curves are recorded across fifteen trials, and are aligned at the moment of triggering button clicks. The black curve is the average measured lateral force and normal force of the fifteen trials, respectively. The red curve in Fig. \ref{fig:force_profile:a} is the target of this rendering. The red line in Fig. \ref{fig:force_profile:b} is the threshold of the normal force that triggers the square-waveform lateral force.}
  \label{fig:force_profile}
\end{figure}

Fig. \ref{fig:force_profile:a} shows the force profile of the finger during button click rendering. Based on the change of the normal force (in Fig. \ref{fig:force_profile:b}), the pressing action starts around 0.26 seconds and lasts 0.44 seconds. The average pressing force is around 900 mN. The command signal is a 160-millisecond rectangular waveform with a 25\% duty cycle and 500 mN peak-to-peak magnitude.

\subsubsection{Experiment 3: Perceptual Evaluation of Rendered Button Clicks}
\begin{figure}[!htb]
\centering
\includegraphics[width = 0.45\textwidth]{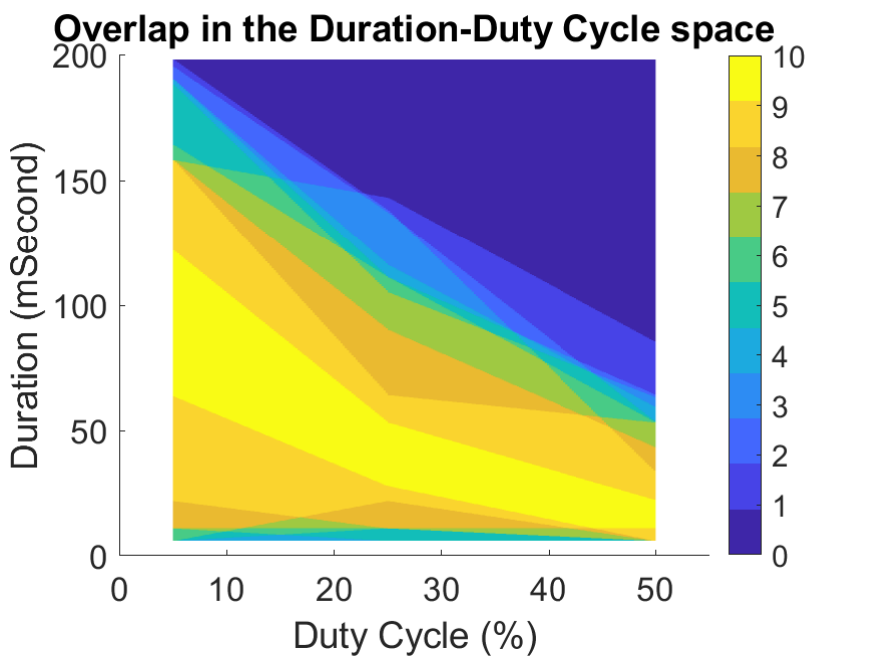}
\caption{The region of the stimulus parameters that are judged to be acceptable button clicks by subjects. The color bar indicates the number of subjects agreein.}
\label{fig_overlap_10Subjects}
\end{figure}

Since the resolution in the duty cycle dimension is low (three levels), we assume that the region of acceptable button clicks for each subject in the duration-duty cycle parameter space can be approximated by the linear connections between the boundaries of the good-button range for each duty cycle (5\%, 25\%, and 50\%). Based on the results of the first question in the first section, the number of subjects who judge that a specific stimulus in the duration-duty cycle parameter space is an acceptable button click is shown in Fig. \ref{fig_overlap_10Subjects}. In this figure, the brightest yellow region indicates all the ten subjects agree that the stimulus is a good button click rendering. The duration ranges of the good rendering are from 70 to 130 milliseconds at the 5\% duty cycle, from 25 to 55 milliseconds at the 25\% duty cycle, and from 10 to 30 milliseconds at the 50\% duty cycle. 

On the contrary, the darkest blue region indicates that all ten subjects judge that the stimulus was not a good button click rendering. In this region, the duration ranges are beyond 200 milliseconds at the 5\% duty cycle, 145 milliseconds at the 25\% duty cycle, and 90 milliseconds at the 50\% duty cycle. 

\begin{figure}[!htb]
\centering
\includegraphics[width = 0.45\textwidth]{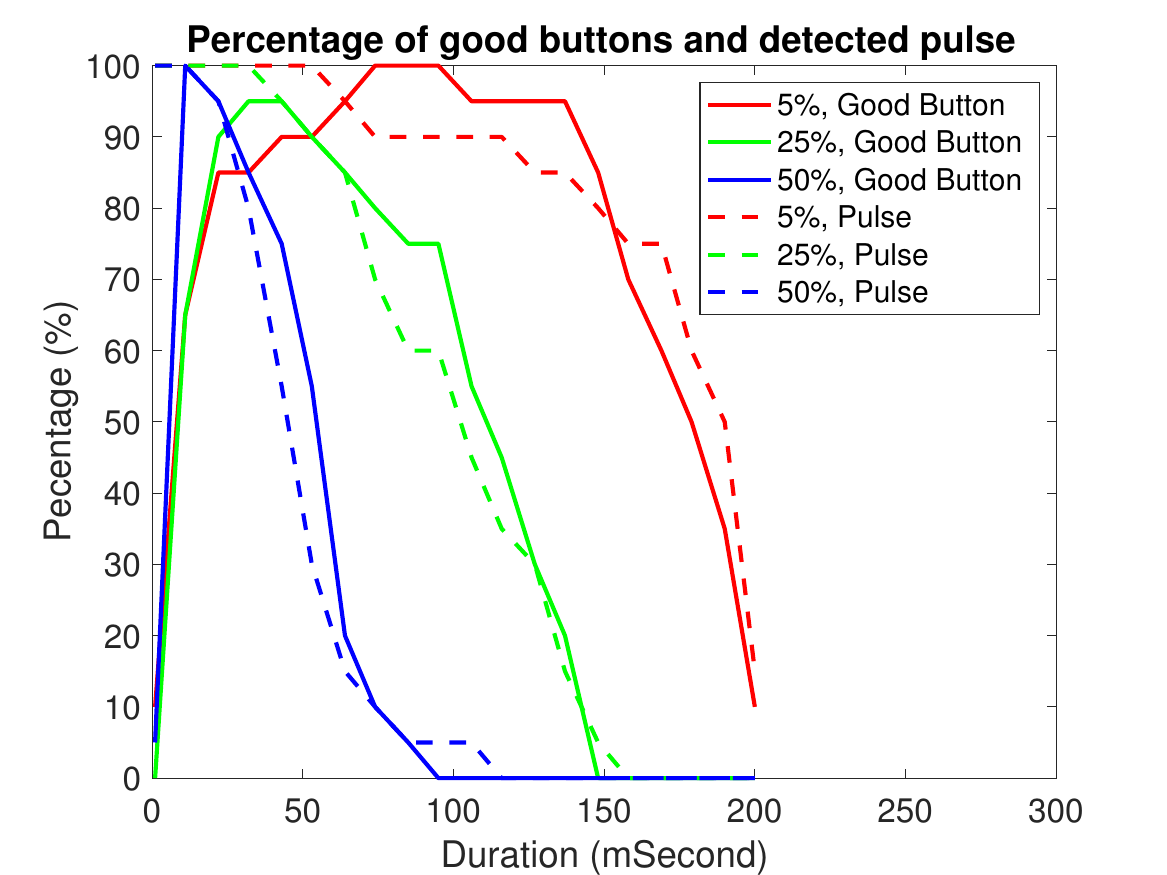}
\caption{Solid lines and dashed lines represent the percentage of subjects reporting a good button click rendering and a detected pulse at the specific duty cycle, respectively.}
\label{fig_percentage_button_pulse}
\end{figure}

Combining the results of the two questions in the first section, Fig. \ref{fig_percentage_button_pulse} shows the percentages of positive responses for both good button click rendering and detecting a pulse (versus an oscillation). The total number of trials for each duration at one duty cycle is 20 (2 trials for each subject). The percentage of good button click rendering at the 50\% duty cycle increases from 5\% to 100\% as the duration increases from 1 to 11 milliseconds, and then decreases to 0\% when the duration is 95 milliseconds. The percentage of good button click rendering at the 5\% and the 25\% duty cycles have similar trends as that at the 50\% duty cycle. When the duration is in the range of 32 to 43 milliseconds, the maximum percentage of the good button click rendering at the 25\% duty cycle is 95\%. When the duration is in the range of 74 to 95 milliseconds, the maximum percentage of the good button click rendering at the 5\% duty cycle is 100\%.

The percentages of a detected pulse are shown as three dashed lines in Fig. \ref{fig_percentage_button_pulse}. For the durations that are less than 11 milliseconds at the 5\% duty cycle, 32 milliseconds at the 25\% duty cycle, or 53 milliseconds at the 50\% duty cycle, all the subjects felt only one pulse in the stimulus. Beyond these duration thresholds, subjects started to feel oscillation.

\begin{figure*}[!htb]
\centering
\includegraphics[width = 0.95\textwidth]{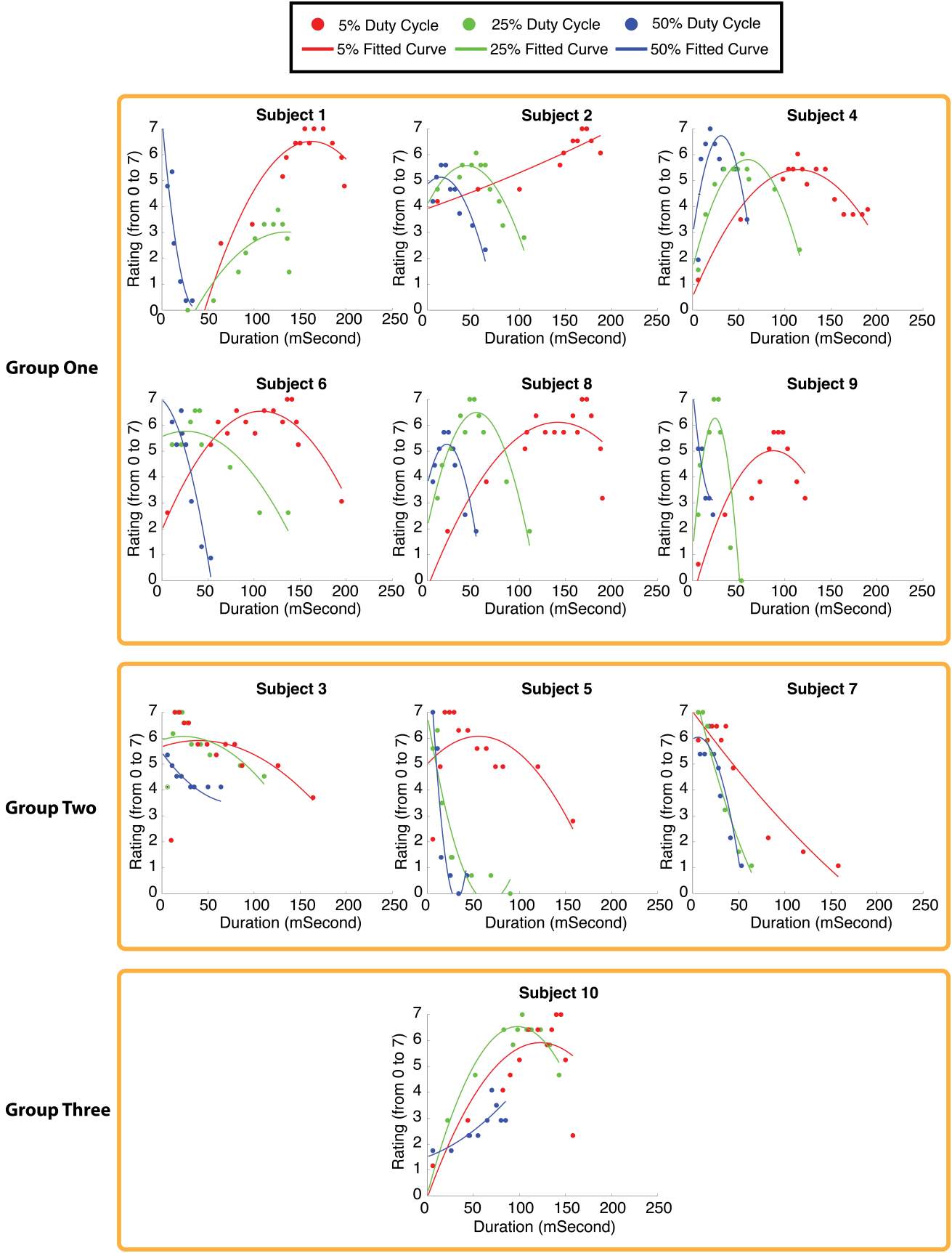}
\caption{Rating of the button click sensation from 0 to 7. Solid lines are the quadratic fitting curves of the rating results. Subject 4, 5, and 9 are graduate students in the haptics group.}
\label{fig_subject_rate}
\end{figure*}

Subject-by-subject ratings of the rendered button click, as determined in the second section of this experiment, are shown in Fig. \ref{fig_subject_rate}. Each subject's rating at three duty cycles is fitted with quadratic equations to help locate the maximum. The results are grouped into three classes based on the relation between subjects' maximum rating and the stimulus duration at the three duty cycles. In group one, the durations of the six subjects' best button clicks increase as the duty cycle decreases. In group two, all three subjects rate short-duration stimuli as best, no matter what the duty cycle is. On the contrary, long-duration stimuli have the highest ratings in group three (a single subject). It is worth noting that some subjects (1, 2, 4, and 9) appear to prefer a specific duty cycle more than they prefer another factor, such as duration. 

\subsection{Richness}
After all the experiments, subjects were invited to report their feeling of the rendered button click and to compare with their own prior experience with buttons. Results are shown in Table \ref{subjectComment}.

\begin{table*}[htp!]
\centering
\caption{Subjects' comments on the rendered button clicks.}
\label{subjectComment}
      \begin{tabular}{cc}
\hline
\hline
Subjects      &  Comments   \\
\hline
 Subject 1 and 2    &  Rendered button clicks are similar to the click on the Magic Trackpad (Apple Inc.)\\

 Subject 3        &  Rendered button clicks are felt like a physical button click with a short travel distance\\
 
 Subject 4, 6 and 8 &  They prefer the stimuli that make them feel a very clear click, rather than oscillations\\

 Subject 5      &  Rendered button clicks are better than the click on the Dell trackpad (Dell Inc.)\\

 Subject 7      &  Rendered button clicks are felt like a click of a mouse\\

 Subject 9      &  Rendered button clicks are similar to a tick\\

 Subject 10     &  Rendered button clicks are felt like virtual clicks that has a travel distance in the normal direction\\
\hline
\hline
\end{tabular}
\end{table*}

\section{Discussion}
\subsection{Localized Control of Lateral Force} \label{discussLocalized}
Fig. \ref{fig:displacement_profile:b} shows that the lateral displacement envelopes of the index finger and the middle finger at 30 kHz are similar. This indicates that the two fingers experience similar vibration amplitudes at 30 kHz due to the vibrating surface even though there is an insulating anodized aluminum foil between the middle finger and the surface.

However, Fig. \ref{fig:displacement_profile:a} shows that the lateral displacement envelope of the index finger at 10 Hz is far higher than that of the middle finger. Since a low-frequency stimulus was used in the button click rendering algorithm, the Meissner corpuscle is expected to dominate the touch sensation. Vibrations of 100 $\mu m$ are enough to stimulate a Meissner corpuscle at 10 Hz \cite{mountcastle1967neural,talbot1968sense,kandel2000principles}. The lateral displacement envelope of the index finger is higher than the human detection threshold of vibration at 10 Hz, which is in turn higher than that of the middle finger. This suggests that only the index finger can feel the active lateral force at 10 Hz. Since the middle finger interacted with an insulating foil made from anodized aluminum, the index finger and the middle finger encountered the same contact characteristics in Experiment 1. We conclude that the low 10 Hz vibration amplitude of the middle finger is due to the absence of the electroadhesion. We should also note that, although the vibration amplitude at the middle finger is about 40 dB lower than at the index finger (Fig. \ref{fig:displacement_profile:a}), it is not zero. This may be due to coupling through the surface or the skeleton of the hand.

These results indicate that the UltraShiver can localize the lateral force on the finger based on localized control of electroadhesion between the finger and the surface. 

\subsection{Force Profile of Button Click Rendering}
Fig. \ref{fig:force_profile:b} shows the normal force profile during pressing. The curves have been aligned at (0.4sec, 600 mN), which is when the lateral force is triggered. The measured square-waveform lateral force applied to the finger (the gray curves in Fig. \ref{fig:force_profile:a}) is closely matched with the command signal, suggesting that the UltraShiver can control the active lateral force and execute the button click rendering algorithm with great precision.

\subsection{Perceptual Experiment}
Fig. \ref{fig_overlap_10Subjects} shows that more than half of stimuli in the duration-duty cycle parameter space are judged by some subjects to be acceptable button clicks. Further, there is an overlap of acceptable button clicks among all the subjects (shown as the brightest yellow region in Fig. \ref{fig_overlap_10Subjects}), despite the fact that subjects made their judgments based on their individual preferences. This suggests that the button click rendering mechanism proposed in the paper has the potential to provide high quality sensations for a large population. 

What are the underlying determinants of a good button click? From a perceptual standpoint, it seems that subjects prefer not to feel an oscillation. This was previously reported by Chen et al. \cite{chen2011design}, it was indicated in our previous work \cite{xu2019localized}, and it is further supported by Experiment 2, the results of which are summarized in Fig. \ref{fig_percentage_button_pulse}. As can be clearly seen, the probability of subjects reporting a good click sensation is closely matched by the probability of reporting a pulse sensation, independent of duty cycle. This, however, begs another question: what characteristics of the force profile determine whether or not the sensation will be experienced as a pulse?

In our experiments, the force profiles are always rectangle waves with a period of high positive force followed by a period of high negative force. This choice was made following preliminary investigations in which the transition from positive to negative was found to contribute to the quality of the click. It is not surprising that a large, rapidly-changing force signal would be salient. Yet, this signal is a constant throughout our experiments, even as the perceived quality of the sensation varies widely. Clearly, additional factors are at play.

Consider, for example, durations of less than 10 milliseconds at the 5\% duty cycle. Here, subjects feel a single pulse, but it is too short to provide a satisfying stimulus. Thus, the pulse duration, not just the maximum slew rate, matters. Longer durations -- 74 to 95 milliseconds -- result in the percept of a single, satisfying click. Even longer durations begin to feel like an oscillation, although it is not clear whether this is due to the perception of back-and-forth movement, due to the perception of multiple transitions (i.e., zero to positive, positive to negative, and negative to zero), or due to some other factor.

\begin{figure}[!htb]
\centering
\includegraphics[width = 0.45\textwidth]{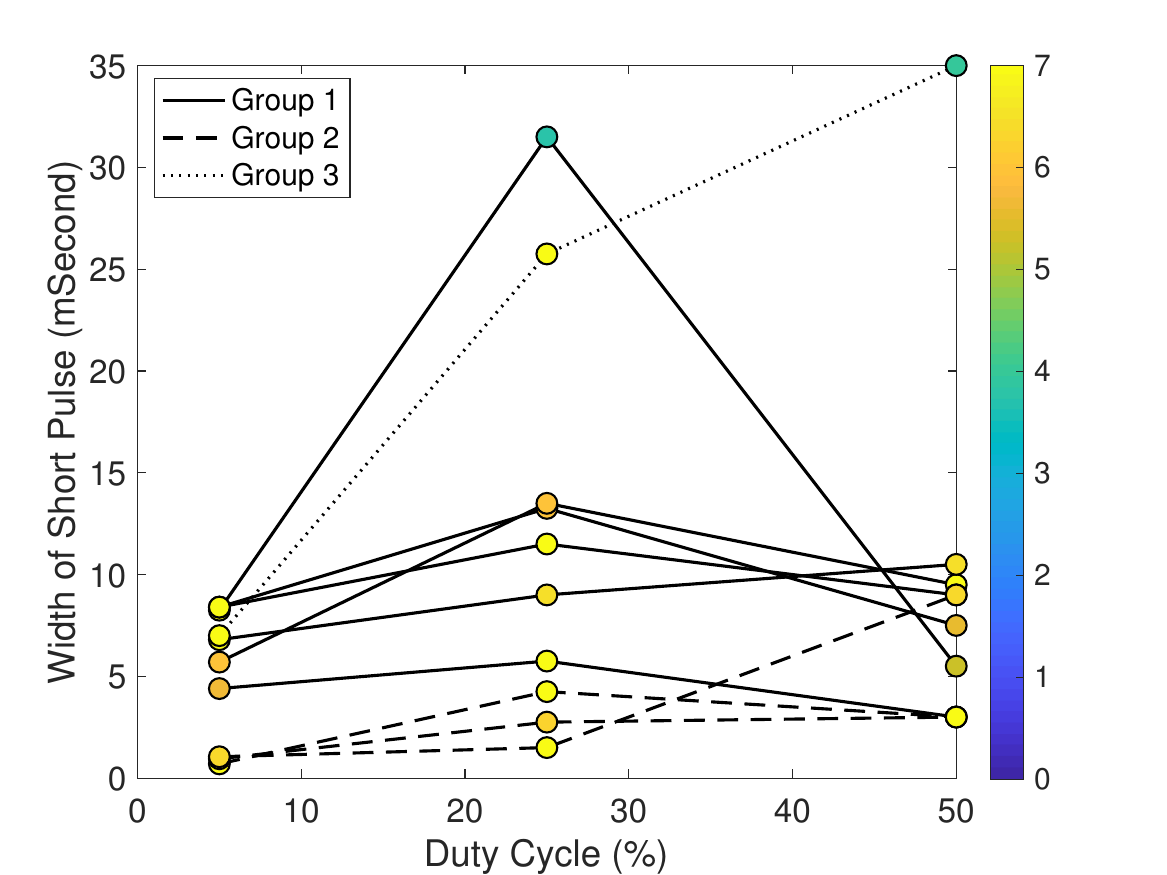}
\caption{Short pulse width of the button click stimulus with the highest rating at three duty cycles. The results of one subject at three duty cycles are connected by the solid line in group one, the dashed line in group two, or the dotted line in group three.}
\label{fig_shortDurationConnect}
\end{figure}

Additional insight comes from the fact that most subjects preferred shorter duration at a longer duty cycle (see Fig. Fig. \ref{fig_overlap_10Subjects}). This suggests that preference may be related to the width of the initial, positive, pulse. Indeed, in \cite{xu2019localized}, we found that the shortest acceptable pulse in the 25\% and 50\% duty cycle cases exhibited similar width ($26.7 \pm 7.6$ milliseconds vs. $26.4 \pm 8.7$ milliseconds). Fig. \ref{fig_shortDurationConnect} shows that, in the current study as well, most subjects preferred a particular value of this initial pulse width, independent of duty cycle. Additional studies will be required to elucidate why this may be. One hypothesis is that subjects want this pulse wide enough to be robustly perceived, and no wider. A weakness of this hypothesis is that in the 5\% and 25\% duty cycle cases, the negative pulse is much wider and should therefore be easily perceived. A second hypothesis is that a minimum amount of time is needed for the applied force to load the tissue of the forefinger, ensuring that the positive-to-negative transition produces a suitably large effect. We examined this by using the LDV to measure the lateral velocity of the fingertip. We found that this velocity saturated at a constant value regardless of pulse duration; thus, the strength of the positive-to-negative transition as measured by fingertip velocity was independent of initial pulse width.  This does not rule out other mechanisms related to initial pulse width, such as the conduction of the impulse through soft tissue or bone to distant mechanoreceptors \cite{libouton2012tactile}.  In future work, we will test these hypotheses with high-bandwidth measurements of finger deformation as well as additional, non-rectangular, force profiles.

It should also be noted that, although all subjects rated selected button clicks quite highly, there was considerable subject-to-subject variability in the specifics. Fig. \ref{fig_subject_rate} presents the subjects' rating of their own acceptable button clicks without normalization. As described earlier, the subjects fall into three groups according to the manner in which preference depends on duty cycle and duration. These same three groups can be seen quite clearly in Fig. \ref{fig_shortDurationConnect}. Subjects prefer either a very short initial pulse (Group 2), a somewhat longer initial pulse (Group 1) or a rather long initial pulse and duration (the individual in Group 3). These differences may be purely perceptual or may relate to fingertip mechanics (e.g., the time to load the tissue) as hypothesized above. This variability will also be a topic of future research.

Overall, these results show that the proposed button click rendering mechanism has the ability to generate a range of realistic button click sensations that can match subjects' individual preferences. In addition to the results in the perceptual experiments, all subjects told the experimenter that they could clearly perceive of click sensations among all the stimuli (shown in Table \ref{subjectComment}). Subjects also reported that some rendered click sensations felt similar to commercial click rendering (Magic Trackpad and Dell Trackpad).

\section{Conclusion}
This study has demonstrated the ability to localize control of an active lateral force on the fingertip and the ability to create a satisfying button click sensation using the Ultrashiver technology. In addition to the results in the perceptual experiments, all subjects told the experimenter that they could clearly perceive of click sensations among all the stimuli (shown in Table \ref{subjectComment}). Subjects also reported that some rendered click sensations felt similar to commercial click rendering (Magic Trackpad and Dell Trackpad). Overall, the ability to produce localized, highly controllable button click rendering, suggests that the technique described here is a promising candidate for touch typing on a flat surface.

\ifCLASSOPTIONcompsoc
  \section*{Acknowledgments}
\else
  \section*{Acknowledgment}
\fi

This material is based upon work supported by the National Science Foundation grants number IIS-1518602.

\bibliographystyle{IEEEtran}
\bibliography{buttonClick}

\begin{thebibliography}{10}
\providecommand{\url}[1]{#1}
\csname url@samestyle\endcsname
\providecommand{\newblock}{\relax}
\providecommand{\bibinfo}[2]{#2}
\providecommand{\BIBentrySTDinterwordspacing}{\spaceskip=0pt\relax}
\providecommand{\BIBentryALTinterwordstretchfactor}{4}
\providecommand{\BIBentryALTinterwordspacing}{\spaceskip=\fontdimen2\font plus
\BIBentryALTinterwordstretchfactor\fontdimen3\font minus
  \fontdimen4\font\relax}
\providecommand{\BIBforeignlanguage}[2]{{%
\expandafter\ifx\csname l@#1\endcsname\relax
\typeout{** WARNING: IEEEtran.bst: No hyphenation pattern has been}%
\typeout{** loaded for the language `#1'. Using the pattern for}%
\typeout{** the default language instead.}%
\else
\language=\csname l@#1\endcsname
\fi
#2}}
\providecommand{\BIBdecl}{\relax}
\BIBdecl

\bibitem{fukumoto2001active}
M.~Fukumoto and T.~Sugimura, ``Active click: tactile feedback for touch
  panels,'' in \emph{CHI'01 Extended Abstracts on Human Factors in Computing
  Systems}.\hskip 1em plus 0.5em minus 0.4em\relax ACM, 2001, pp. 121--122.

\bibitem{chen2011design}
H.-Y. Chen, J.~Park, S.~Dai, and H.~Z. Tan, ``Design and evaluation of
  identifiable key-click signals for mobile devices,'' \emph{IEEE Transactions
  on Haptics}, vol.~4, no.~4, pp. 229--241, 2011.

\bibitem{weir2004haptic}
D.~W. Weir, M.~Peshkin, J.~E. Colgate, P.~Buttolo, J.~Rankin, and M.~Johnston,
  ``The haptic profile: capturing the feel of switches,'' in \emph{Haptic
  Interfaces for Virtual Environment and Teleoperator Systems, 2004.
  HAPTICS'04. Proceedings. 12th International Symposium on}.\hskip 1em plus
  0.5em minus 0.4em\relax IEEE, 2004, pp. 186--193.

\bibitem{zoller2012novel}
I.~Zoller, P.~Lotz, and T.~A. Kern, ``Novel thin electromagnetic system for
  creating pushbutton feedback in automotive applications,'' in
  \emph{International Conference on Human Haptic Sensing and Touch Enabled
  Computer Applications}.\hskip 1em plus 0.5em minus 0.4em\relax Springer,
  2012, pp. 637--645.

\bibitem{kessler2015haptic}
P.~Kessler, D.~C. Patel, J.~A. Harley, B.~W. Degner, N.~A. Rundle, P.~K.
  Augenbergs, N.~Lubinski, K.~L. Staton, and O.~S. Leung, ``Haptic
  electromagnetic actuator,'' Apr.~23 2015, uS Patent App. 14/404,156.

\bibitem{gueorguiev2018travelling}
D.~Gueorguiev, A.~Kaci, M.~Amberg, F.~Giraud, and B.~Lemaire-Semail,
  ``Travelling ultrasonic wave enhances keyclick sensation,'' in
  \emph{International Conference on Human Haptic Sensing and Touch Enabled
  Computer Applications}.\hskip 1em plus 0.5em minus 0.4em\relax Springer,
  2018, pp. 302--312.

\bibitem{tashiro2009realization}
K.~Tashiro, Y.~Shiokawa, T.~Aono, and T.~Maeno, ``Realization of button click
  feeling by use of ultrasonic vibration and force feedback,'' in
  \emph{EuroHaptics conference, 2009 and Symposium on Haptic Interfaces for
  Virtual Environment and Teleoperator Systems. World Haptics 2009. Third
  Joint}.\hskip 1em plus 0.5em minus 0.4em\relax IEEE, 2009, pp. 1--6.

\bibitem{monnoyrer2018perception}
J.~Monnoyrer, E.~Diaz, C.~Bourdin, and M.~Wiertlewski, ``Perception of
  ultrasonic switches involves large discontinuity of the mechanical
  impedance,'' \emph{IEEE transactions on haptics}, 2018.

\bibitem{hudin2015localized}
C.~Hudin, J.~Lozada, and V.~Hayward, ``Localized tactile feedback on a
  transparent surface through time-reversal wave focusing,'' \emph{Ieee
  transactions on haptics}, no.~2, pp. 188--198, 2015.

\bibitem{hudin2017local}
C.~Hudin, ``Local friction modulation using non-radiating ultrasonic
  vibrations,'' in \emph{World Haptics Conference (WHC), 2017 IEEE}.\hskip 1em
  plus 0.5em minus 0.4em\relax IEEE, 2017, pp. 19--24.

\bibitem{xu2019ultrashiver}
H.~Xu, M.~A. Peshkin, and E.~Colgate, ``Ultrashiver: Lateral force feedback on
  a bare fingertip via ultrasonic oscillation and electroadhesion,'' \emph{IEEE
  transactions on haptics}, 2019.

\bibitem{xu2019localized}
H.~Xu, R.~L. Klatzky, M.~A. Peshkin, and J.~E. Colgate, ``Localized rendering
  of button click sensation via active lateral force feedback,'' in \emph{2019
  IEEE World Haptics Conference (WHC)}.\hskip 1em plus 0.5em minus 0.4em\relax
  IEEE, 2019, pp. 509--514.

\bibitem{peshkin2014haptic}
M.~A. Peshkin and J.~E. Colgate, ``Haptic display with simultaneous sensing and
  actuation,'' US Patent App. 14/306,842, Dec.~25 2014.

\bibitem{colgate2017touch}
J.~E. Colgate and M.~A. Peshkin, ``Touch interface device having an
  electrostatic multitouch surface and method for controlling the device,'' US
  Patent 9,733,746, Aug.~15 2017.

\bibitem{shultz2018application}
C.~Shultz, M.~Peshkin, and J.~E. Colgate, ``The application of tactile,
  audible, and ultrasonic forces to human fingertips using broadband
  electroadhesion,'' \emph{IEEE transactions on haptics}, vol.~11, no.~2, pp.
  279--290, 2018.

\bibitem{mountcastle1967neural}
V.~B. Mountcastle, W.~H. Talbot, I.~Darian-Smith, and H.~H. Kornhuber, ``Neural
  basis of the sense of flutter-vibration,'' \emph{Science}, vol. 155, no.
  3762, pp. 597--600, 1967.

\bibitem{talbot1968sense}
W.~H. Talbot, I.~Darian-Smith, H.~H. Kornhuber, and V.~B. Mountcastle, ``The
  sense of flutter-vibration: comparison of the human capacity with response
  patterns of mechanoreceptive afferents from the monkey hand.'' \emph{Journal
  of neurophysiology}, vol.~31, no.~2, pp. 301--334, 1968.

\bibitem{kandel2000principles}
E.~R. Kandel, J.~H. Schwartz, T.~M. Jessell, D.~of~Biochemistry, M.~B.~T.
  Jessell, S.~Siegelbaum, and A.~Hudspeth, \emph{Principles of neural
  science}.\hskip 1em plus 0.5em minus 0.4em\relax McGraw-hill New York, 2000,
  vol.~4.

\bibitem{biggs2002tangential}
J.~Biggs and M.~A. Srinivasan, ``Tangential versus normal displacements of
  skin: Relative effectiveness for producing tactile sensations,'' in
  \emph{Proceedings 10th Symposium on Haptic Interfaces for Virtual Environment
  and Teleoperator Systems. HAPTICS 2002}.\hskip 1em plus 0.5em minus
  0.4em\relax IEEE, 2002, pp. 121--128.

\bibitem{libouton2012tactile}
X.~Libouton, O.~Barbier, Y.~Berger, L.~Plaghki, and J.-L. Thonnard, ``Tactile
  roughness discrimination of the finger pad relies primarily on vibration
  sensitive afferents not necessarily located in the hand,'' \emph{Behavioural
  brain research}, vol. 229, no.~1, pp. 273--279, 2012.

\end{thebibliography}

\begin{IEEEbiography}[{\includegraphics[width=1in,height=1.25in,clip,keepaspectratio]{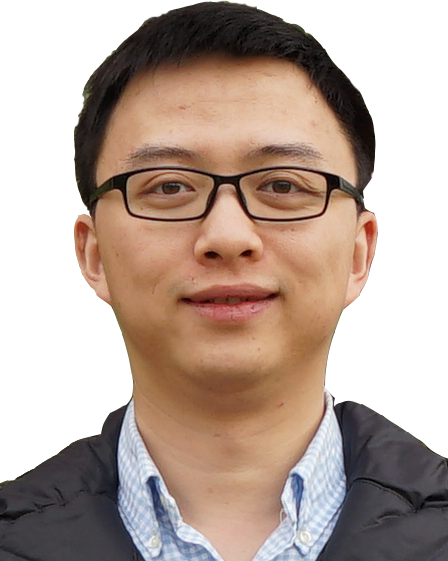}}]{Heng Xu}
is a Ph.D. candidate in the Department of Mechanical Engineering, Northwestern University, Evanston, IL, USA. His research interests include surface haptics, electrotactile stimulation, and tactile sensory feedback. He is a member of the IEEE.
\end{IEEEbiography}

\begin{IEEEbiography}[{\includegraphics[width=1in,height=1.25in,clip,keepaspectratio]{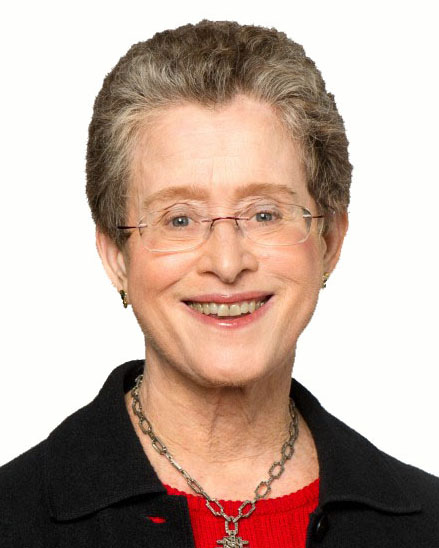}}]{Roberta L. Klatzky}
received the Ph.D. degree in cognitive psychology from Stanford University. She is currently the Charles J. Queenan, Jr. professor of psychology and human–computer interaction at Carnegie Mellon University, Pittsburgh, PA, USA. Her research interests include human perception and cognition, with special emphasis on spatial cognition and haptic perception. She is a Fellow of the IEEE.
\end{IEEEbiography}

\begin{IEEEbiography}[{\includegraphics[width=1in,height=1.25in,clip,keepaspectratio]{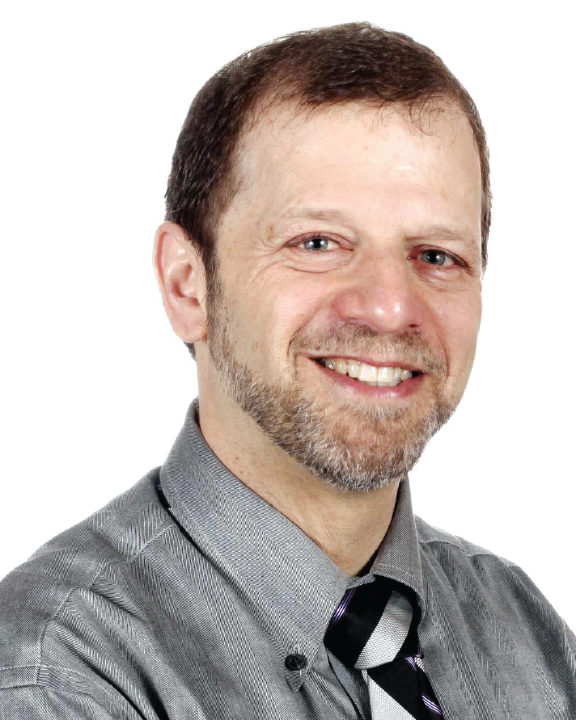}}]{Michael A. Peshkin}
is the Bette and Neison Harris professor of teaching excellence in the Department of Mechanical Engineering, Northwestern University, Evanston, Illinois. His research is in haptics, robotics, human-machine interaction, and rehabilitation robotics. He has co-founded four start-up companies: Mako Surgical, Cobotics, HDT Robotics, and Tanvas. He is a fellow of the National Academy of Inventors, and (with J. E. Colgate) an inventor of cobots. He is a senior member of the IEEE.
\end{IEEEbiography}

\begin{IEEEbiography}[{\includegraphics[width=1in,height=1.25in,clip,keepaspectratio]{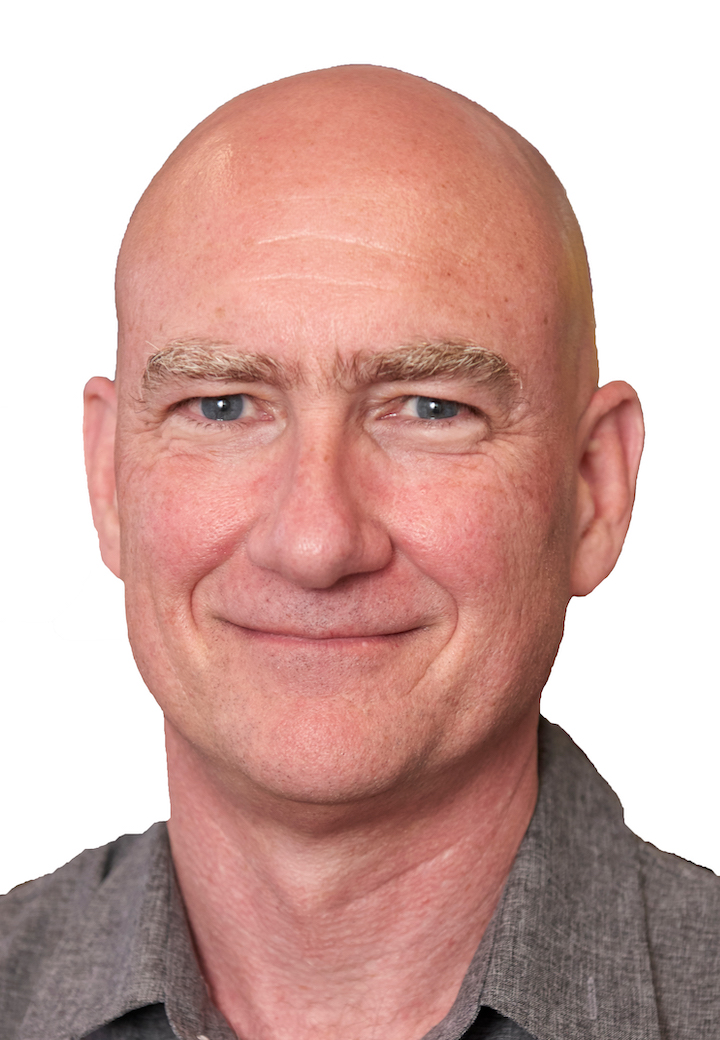}}]{J. Edward Colgate}
is the Breed University Design professor at Northwestern University. He is known for his work in haptics and human-robot collaboration. He served as an associate editor of the IEEE Transactions on Robotics and Automation, and he was the founding editor-in-chief of the IEEE Transactions on Haptics. He was one of the founding codirectors of the Segal Design Institute, Northwestern University. He has founded three start-up companies the most recent of which, Tanvas Inc., is commercializing
an innovative haptic touch screen. He is a fellow of the IEEE and the National Academy of Inventors.
\end{IEEEbiography}

\end{document}